\documentclass{ws-procs961x669}          

\usepackage{mathrsfs}

\def\figsubcap#1{\par\noindent\centering\footnotesize(#1)}
\newcommand*{\diff}{\mathop{}\!\mathrm{d}}
\newcommand*{\f}{\mathrm{f}}
\renewcommand*{\v}{\mathrm{v}}
\newcommand*{\e}{\mathrm{e}}
\newcommand*{\n}{\mathrm{n}}
\newcommand*{\F}{\mathrm{F}}
\newcommand*{\E}{\mathrm{E}}
\newcommand*{\SL}{\mathrm{SL}(2,\mathbb{C})}
\newcommand*{\SU}{\mathrm{SU}(2)}

\begin{document}
\title{A spin foam framework for the black-to-white hole transition}

\author{Farshid Soltani}

\address{Department of Physics and Astronomy, The University of Western Ontario,\\
London, Ontario N6A 3K7, Canada\\
E-mail: fsoltan2@uwo.ca}

\begin{abstract}
Black holes formation and evolution have been extensively studied at the classical level. However, not much is known regarding the end of their lives, a phase that requires to consider the quantum nature of the gravitational field. A black-to-white hole transition can capture the physics of this phenomenon, in particular the physics of the residual small black holes at the end of the Hawking evaporation. This work shows how the spin foam formalism is able to describe this non-perturbative phenomenon. A thorough examination of the black hole spacetime region in which quantum effects cannot be neglected indicates that the scenario in which the black hole geometry undergoes a quantum transition in a white hole geometry is natural and conservative. This quantum transition is then studied using the spin foam formalism and the resulting transition amplitude is explicitly computed. 
\end{abstract}

\keywords{Black holes, white holes, black-to-white hole transition, loop quantum gravity, spin foam formalism.}

\bodymatter

\section{Introduction}

The exterior region of a black hole is described extraordinarily well by general relativity. Its interior region, on the other hand, is not and it is thus not known what happens inside a black hole. The reason for this breakdown of predictability is the presence of a spacetime singularity in the interior of a black hole: since the quantum nature of the gravitational field cannot
be neglected in the vicinity of a spacetime singularity, the classical theory cannot be trusted in this region. Although effective black hole models exhibiting a non-singular interior have been extensively studied in the literature, a non-singular black hole interior consisting of a black hole geometry that undergoes a quantum transition in a white hole geometry was proposed for the first time in Ref.~\citenum{Haggard_2015_fireworks}.

A different open question concerning black holes is what happens at the end of their life. Working in semiclassical gravity, Hawking\cite{article:Hawking_BH_explosion} famously showed that black holes evaporate and are thus not eternal. However, since the quantum nature of the gravitational field near the horizon can no longer be neglected when the horizon reaches Planckian size (or possibly even before then\cite{Haggard_2015_fireworks}), the end of the evaporation process of a black hole is a quantum gravity phenomenon. A very natural and conservative scenario for the end of the life of a black hole is the black-to-white hole transition.

The aim of this work is to report in a concise and coherent fashion the results obtained in Refs.~\citenum{article:DAmbrosio_2021} and~\citenum{article:soltani2021}. The black hole spacetime region in which quantum effects cannot be neglected is analyzed in detail and it is shown to be actually composed of three physically independent subregions, with one of them being a region surrounding the black hole horizon at the end of the evaporation process. As a consequence, the last stage of the life of a black hole can be studied focusing solely on this region. Independent analyses of the three separate quantum regions consistently point towards a black-to-white hole transition. The last stage of the life of a black hole in this scenario, that is the quantum region where the black hole horizon undergoes a quantum transition in a white hole horizon, is then studied using the spin foam formalism\cite{book:Rovelli_Vidotto_CovariantLQG,article:Perez_2013} (also known as covariant loop quantum gravity) and a concrete spin foam framework for the black-to-white hole transition is developed.

The discussion is here limited to the case of a Schwarzschild black hole. See however Ref.~\citenum{rignonbret2021black} for a generalization of the black-to-white hole geometry to the case of a charged black hole.

\section{The three quantum regions of a black hole spacetime}

\begin{figure}[b]
\centering
\includegraphics[scale=0.25]{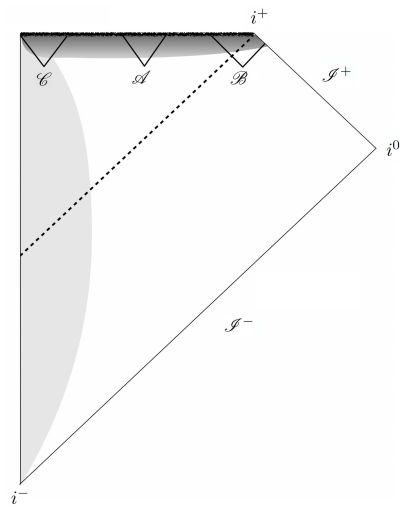}
\caption{Conformal diagram of the spacetime describing the formation of a black hole by gravitationally collapsed matter and its subsequent evaporation.}
\label{fig:ABC}
\end{figure}

The conformal diagram of the spacetime describing the formation of a black hole by gravitationally collapsed matter and its subsequent evaporation is reported in Fig.~\ref{fig:ABC}. The light grey region represents the interior of the collapsing matter, the dashed line represents the apparent horizon of the black hole and the dark grey region represents the spacetime region where the quantum nature of the gravitational field cannot be neglected and where consequently the classical (or semiclassical) theory can no longer be trusted.

The quantum region can be divided\cite{article:Bianchi_etal_2018_remnants} in three different subregions: region $\mathscr{C}$, where the collapsing matter reaches Planckian density; region $\mathscr B$, where the horizon reaches Planckian size at the end of the evaporation process; region $\mathscr A$, where the curvature reaches Planckian scale but the spacetime events belong neither to region $\mathscr B$ nor to region $\mathscr C$. In order to prove that these subregions are physically $\,\!$independent from each other, it is sufficient to show $\,\!$that their spatial separation is $\,\!$considerable. In the classical theory the principle of $\,\!$locality assures that two events $\,\!$whose separation is spacelike and significant cannot be $\,\!$causally connected. Furthermore, although the $\,\!$quantum theory may alter $\,\!$the causal structure of spacetime, quantum fluctuations cannot $\,\!$causally connect widely spacelike-separated events.

A rough estimate of the $\,\!$distance $d_{\mathscr BC}$ between regions $\mathscr B$ and $\mathscr C$ is\cite{article:DAmbrosio_2021}
\begin{equation}
d_{\mathscr BC} \sim \:\ell_{\mathrm Pl} \: \bigg( \frac{m_0}{m_{\mathrm Pl}} \bigg)^{\frac{10}{3}}\, ,
\label{eq:distance_BC}
\end{equation}
where $m_0$ is the initial mass $\,\!$of the black hole before the start of the evaporation process and $\ell_{\mathrm Pl}$ and $m_{\mathrm Pl}$ are respectively the Planck length and the Planck mass. For a stellar $\,\!$black hole this distance is
\begin{equation}
d_{\mathscr BC} \big(m_0 = M_{\odot}\big)\sim 10^{75} \:\,\mathrm{light\:\, years}\, .
\label{eq:distance_BC_stellarmass}
\end{equation}
Regions $\mathscr B$ and $\mathscr C$ are thus spacelike $\,\!$separated and considerably distant from $\,\!$each other. It follows that regions $\mathscr B$ and $\mathscr C$ are physically $\,\!$independent from each other and they $\,\!$can thus be studied separately. Furthermore, since region $\mathscr A$ contains spacetime $\,\!$events that are physically independent from $\,\!$region $\mathscr B$ as well as spacetime events that are physically $\,\!$independent from region $\mathscr C$, the physics of region $\mathscr A$ must be independent $\,\!$from the physics of both region $\mathscr B$ and region $\mathscr C$. However, since $\,\!$region $\mathscr A$ contains spacetime events that $\,\!$are causally connected to region $\mathscr B$ as well as spacetime events $\,\!$that are causally connected to region $\mathscr C$, the physics of both $\,\!$region $\mathscr B$ and region $\mathscr C$ depend on the physics of $\,\!$region $\mathscr A$. These three quantum regions will be now analyzed separately.

\subsection{Region $\mathscr{A}$}

Since the physics of $\,\!$region $\mathscr A$ does not depend on the quantum gravity regime of the collapsing matter (region $\mathscr C$) or on the last stage of the life of the black hole (region $\mathscr B$), it can be studied in the context of an eternal (Schwarzschild) black hole.

The effort to understand the physics of region $\mathscr A$ has led to the development of several effective models (see Refs.~\citenum{article:Singh_Corichi_2016,article:Ashtekar_Olmedo_2018,article:gambini_2020} and references therein) that describe the internal region of an eternal black hole using techniques developed in the framework of loop quantum cosmology. Although the specifics of these models are different, they all exhibit a regular interior region where the trapped region of the black hole makes a smooth transition in an anti-trapped region bounded by a future horizon describing the interior of a white hole. This result, besides supporting the conjecture that classical curvature singularities are not a true physical prediction of the theory but rather an indication that the classical theory can no longer be trusted, is the first evidence suggesting a black-to-white hole transition.

Interestingly, a hint of the same result can be found also at the classical level. It can indeed be shown\cite{Peeters_1995,D_Ambrosio_2018} that, using specific coordinate systems, the geodesics in the interior region of a black hole can be naturally continued across the curvature singularity into the interior region of a white hole. The resulting geometry is still singular, but now it is geodesically complete and it can be interpreted as the classical limit of the quantum geometry of the effective models.

\subsection{Region $\mathscr{C}$}

The physics of region $\mathscr C$ is independent from the physics of region $\mathscr B$ and it can thus be studied neglecting the evaporation process of the black hole. However, it is not independent from the physics of region $\mathscr A$ and its analysis must be consistent with the scenario emerging from the investigation of region $\mathscr A$.

The study of the classical physics of the collapsing matter is a hard task. The analysis of its quantum gravity regime is even trickier. In Ref.~\citenum{Planck_Stars} it was hypothesized that, in analogy with the cosmological singularity resolution in loop quantum cosmology,\cite{Ashtekar_2006_BigBang} the collapsing matter bounces due to a quantum-gravitational repulsion effect. This possibility, that will be assumed to accurately represent the physics of region $\mathscr C$ in the following, is consistent with the physics of region $\mathscr A$. A qualitative conformal diagram of the spacetime emerging from this partial analysis (region $\mathscr B$ still needs to be discussed) can be found in Fig.~\ref{fig:B2W}(a).

This scenario is further corroborated by several independent quantum descriptions of the phenomenon (see e.g. Refs.~\citenum{hajicek_kiefer_2001,Kelly_2020,Piechocki_2020,Munch_2021}). Although these models use different techniques and focus on different aspects of the phenomenon, they all predict that the collapsing matter undergoes a bounce. This is a strong indication of the general validity of the scenario.

\subsection{Region $\mathscr{B}$}

\begin{figure}[t]%
\begin{center}
\parbox{2.1in}{\includegraphics[scale=0.3]{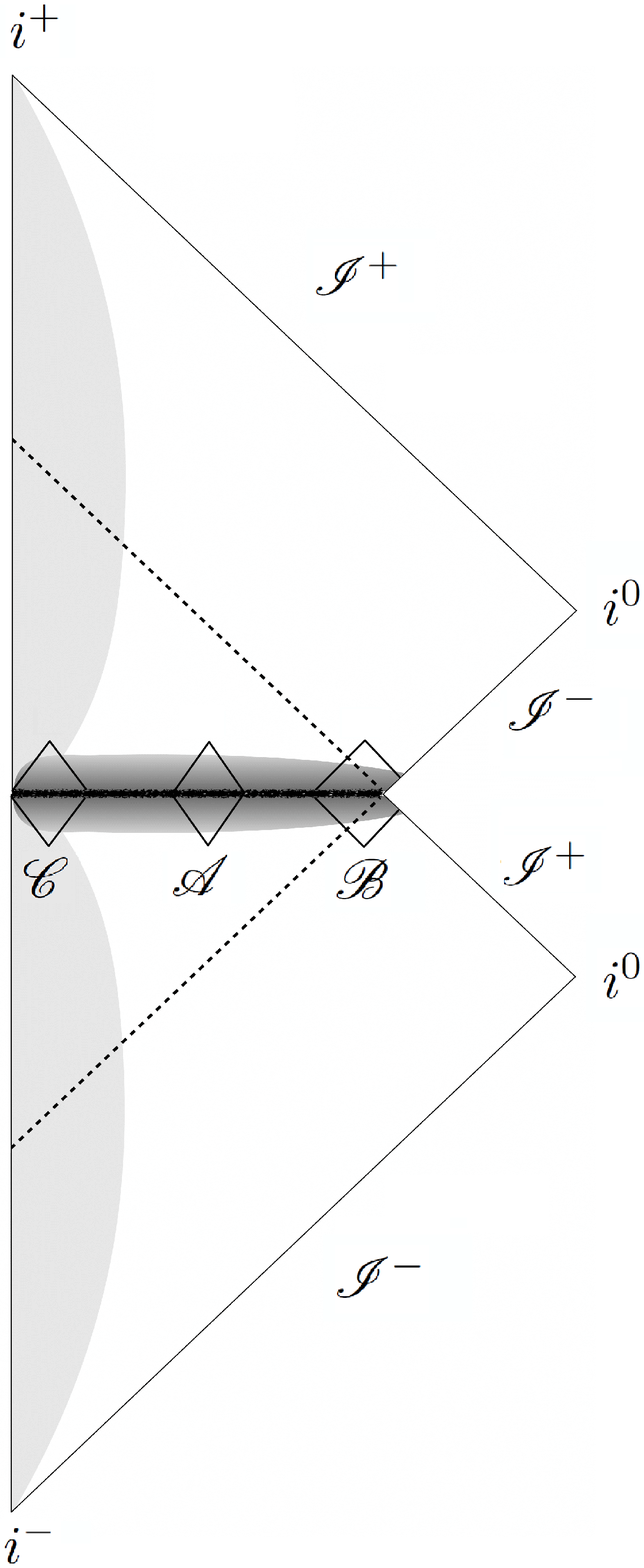}
\figsubcap{a}}
\hspace*{25pt}
\parbox{2.1in}{\includegraphics[scale=0.295]{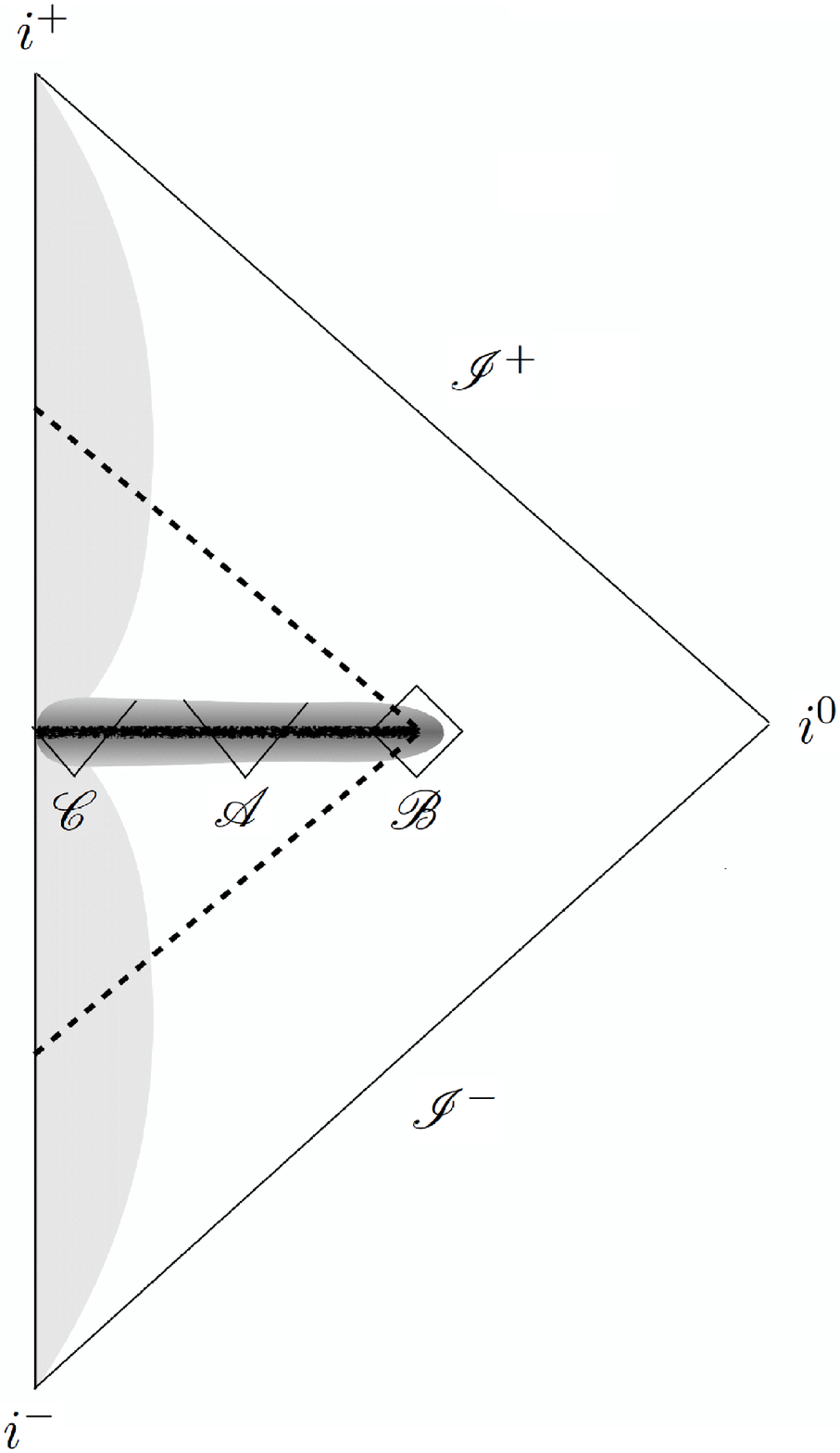}
\figsubcap{b}}
\caption{(a) Qualitative conformal diagram of the black hole spacetime emerging from the analysis of regions $\mathscr A$ and $\mathscr C$. (b) Conformal diagram of the spacetime describing the black-to-white hole transition.}
\label{fig:B2W}
\end{center}
\end{figure}

The conformal diagram in Fig.~\ref{fig:B2W}(a) represents the black hole spacetime that emerges taking into account the quantum physics of regions $\mathscr A$ and $\mathscr C$. It is however immediate to see that this spacetime does not represent properly the physics of region $\mathscr{B}$. While the black hole evaporation process takes a finite amount of time to shrink the horizon to the Planck scale, in Fig.~\ref{fig:B2W}(a) region $\mathscr{B}$ is reached only asymptotically. The physics of region $\mathscr{B}$ thus need to be properly modified whilst remaining consistent with the scenario emerging from regions $\mathscr A$ and $\mathscr C$.

A scenario that is often considered for the end of the life of a black hole is the complete evaporation of the black hole. Having spent time investigating regions $\mathscr A$ and $\mathscr C$ separately, it is now easy to see that a complete evaporation of the black hole, besides being an ad hoc assumption with no foundation in any quantum gravity model, is hardly consistent with the global picture of the spacetime in Fig.~\ref{fig:B2W}(a). The most natural and conservative scenario for the end of the life of a black hole consistent with the physics of region $\mathscr A$ and region $\mathscr C$ is a quantum transition of the black hole horizon in a white hole horizon. The conformal diagram of the spacetime describing the complete black-to-white hole transition can be found in Fig.~\ref{fig:B2W}(b). The purpose of this work is to complete the analysis of region $\mathscr B$ by investigating its quantum physics using the spin foam approach.

The conformal diagram in Fig.~\ref{fig:B2W}(b) describes a physical phenomenon only if there exists a classical metric that satisfies the Einstein field equations and that covers the whole diagram except for the quantum region. This metric exists and it has been explicitly constructed in Refs.~\citenum{Haggard_2015_fireworks,Pierre_2018,Pierre_2019_B2Wevaporating}. This is an extraordinary result from the point of view of the classical theory and it provides yet another strong evidence in favor of the black-to-white hole transition scenario.

\section{The black-to-white hole transition}

In order to investigate in detail the physics of region $\mathscr B$ of the black-to-white hole spacetime it is necessary to specify its boundary $\Sigma:= \partial \mathscr B$ and to compute the classical geometry that the black-to-white hole metric induces on it. This boundary geometry represents the outcome of the quantum transition taking place in region $\mathscr B$ and it thus uniquely defines the quantum boundary state for the transition. The boundary $\Sigma$ can be chosen freely as long as it bounds the entirety of the quantum subregion.

As a first $\,\!$approximation, the presence of the $\,\!$Hawking radiation near $\,\!$region $\mathscr B$ is neglected. Its inclusion is left for future work. The metric $\,\!$around region $\mathscr B$ is thus taken to be the Schwarzschild metric up to quantum corrections from region $\mathscr A$. The main quantum correction that the physics of region $\mathscr A$ induces on the boundary between regions $\mathscr A$ and $\mathscr B$ is the absence of the classical singularity. The black hole interior $\,\!$can be foliated with $\,\!$surfaces of topology $S^2 \times \mathbb{R}$. If one angular $\,\!$dimension is $\,\!$suppressed, these surfaces can be seen as long cylinders $\,\!$of different radii and $\,\!$heights. Closer is the singularity, smaller is the radius of the cylinder. In the classical theory the foliation ends at the singularity, where the cylinder has a null radius. In the quantum theory the cylinder radius shrinks (black hole geometry) until it reaches a minimum value $r_*$ (smooth transition from black to white hole geometry) and then starts to increase again (white hole geometry). The presence of this minimum radius $r_*$ in the effective geometry of region $\mathscr A$ has a significant impact on the physics of region $\mathscr B$.

\begin{figure}[b]
\centering
\includegraphics[scale=0.2]{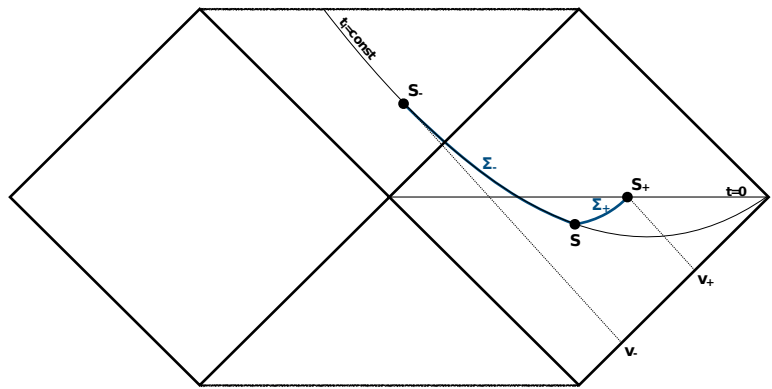}
\caption{The past portion $\Sigma^p=\Sigma^p_+ \cup \Sigma^p_-$ of the boundary $\Sigma$.}
\label{fig:boundary}
\end{figure}

\begin{figure}[t]
\centering
\includegraphics[scale=0.3]{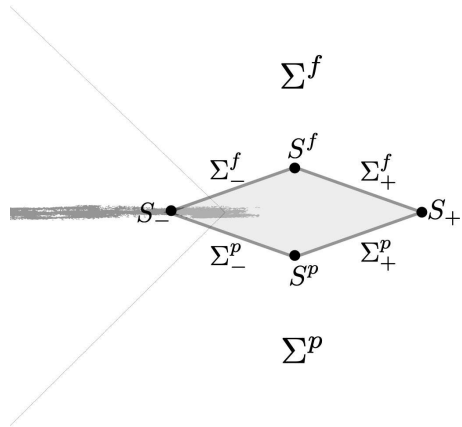}
\caption{Portion of the conformal diagram of the black-to-white hole transition depicting region $\mathscr B$ and its boundary $\Sigma$.}
\label{fig:sigma}
\end{figure}

Having neglected Hawking $\,\!$radiation, the physics of $\,\!$region $\mathscr B$ must be invariant $\,\!$under time-reversal. Accordingly, $\,\!$the boundary $\Sigma$ can be $\,\!$decomposed as $\Sigma~=~\Sigma^p~\cup~\Sigma^f$, where the past $\,\!$surface $\Sigma^p$ and the future $\,\!$surface $\Sigma^f$ are equal up to $\,\!$time reflection. This$\,\!$ means that to completely $\,\!$define $\Sigma$ it is sufficient to only $\,\!$specify $\Sigma^p$. A convenient choice $\,\!$of $\Sigma^p$ is reported $\,\!$in Fig.~\ref{fig:boundary} and it $\,\!$can be constructed as follows. Let $S_+$ $\,\!$and $S_-$ be the $\,\!$points in the conformal diagram of $\,\!$Schwarzschild spacetime defined by the ingoing $\,\!$Eddington-Finkelstein coordinates $S_+=(v_+ , r_+)$ and $S_-=(v_-, r_-)$. The $\,\!$point $S_+$ is taken to be on the surface of constant Schwarzschild $\,\!$time $t=0$, thus fixing the value of $v_+$ in terms of $r_+$ (or viceversa), and $\,\!$the point $S_-$ is taken to be on the surface of $\,\!$constant Schwarzschild radius $r=r_*$, thus fixing the value of $r_-$ to be the minimum radius $r_*$ characterizing the effective geometry of region $\mathscr A$. The values of $r_+$ and $v_-$ can be chosen freely as long as they define a boundary $\Sigma$ that bounds the entirety of the quantum subregion. Let $\,\!$then $\Sigma^{p}_-$ be the surface of $\,\!$constant Lemaitre time coordinate
\begin{equation}
t_L := t + 2 \sqrt{2mr} \,\! + 2m \, \log \bigg| \frac{\,\!\sqrt{r/2m} - 1}{\sqrt{r/ 2m}+ \,\! 1}\bigg|
\label{eq:lemaitre}
\end{equation}
passing $\,\!$by $S_-$, where $m$ is the mass of the black hole at the moment in which the quantum transition of the horizon takes place, and $\Sigma^{p}_+$ be the $\,\!$surface of $\,\!$equation
\begin{equation}
v-\beta \, r = \mathrm{const}\, ,
\label{eq:sigma+}
\end{equation}
where $\beta$ is an arbitrary constant in $\mathbb{R}$, passing $\,\!$by $S_+$. Given the point $S^p$ of their intersection, the past $\,\!$boundary $\Sigma^p$ is defined to be the $\,\!$union of the portion of the surface $\Sigma^{p}_-$ bounded $\,\!$by $S_-$ and $S^p$ and the portion of the $\,\!$surface $\Sigma^{p}_+$ bounded by $S_+$ and $S^p$. Requiring the $\,\!$normal to $\Sigma^p$ to be continuous at $S^p$ uniquely $\,\!$fixes the value of $\beta$. The $\,\!$surface $\Sigma^f$ is defined as $\,\!$the time-reversal of $\Sigma^p$. The portion of the $\,\!$conformal diagram of the black-to-white hole transition depicting region $\mathscr B$ and its boundary $\Sigma$ can be found in Fig.~\ref{fig:sigma}.

The intrinsic and the extrinsic geometry of $\Sigma^{p}_+$ and $\Sigma^{p}_-$ can be straightforwardly computed from their definition. The line element $\diff s^2_+$ on $\Sigma^{p}_+$ reads
\begin{equation}
\diff s^2_+ = \beta \, \bigg( 2- \beta \Big( 1 - \frac{2m}{r} \Big) \bigg) \diff r^2 + r^2 \diff \Omega^2
\end{equation}
and the line element $\diff s^2_-$ on $\Sigma^{p}_-$ reads
\begin{equation}
\diff s^2_- =  \diff r^2 + r^2 \diff \Omega^2\, ,
\end{equation}
where $\diff \Omega^2$ is the line element of the two-sphere. Let $k^{\pm}_{ij}$ be the extrinsic curvature tensor of $\Sigma^{p}_\pm$. Then it can be shown that
\begin{equation}
k_+ := k^{+}_{ij} \diff x^i \diff x^j =  \frac{ m \beta^{3 /2} (r (3- \beta ) +2m \beta) }{ \sqrt{ r^5 ( r(2 -\beta ) + 2 m \beta) } }\: \diff r^2 - \frac{ r (1 -\beta) +2 m \beta}{ \sqrt{ \beta( 2- (1- 2m/  r) \beta ) } } \: \diff \Omega^2 
\label{eq:k1}
\end{equation}
and 
\begin{equation}
k_- := k^{-}_{ij} \diff x^i \diff x^j = \frac{m}{2 r^3} \diff r^2 - \sqrt{2mr} \diff \Omega^2\, ,
\label{eq:k2}
\end{equation}
where $x^i$, $i=1,2,3$ , are coordinates on $\Sigma^{p}$.

The geometry of $\Sigma$ is thus $\,\!$completely determined by four $\,\!$parameters: the mass $m$ of the black $\,\!$hole at the moment $\,\!$in which the quantum transition of the horizon takes place; the external $\,\!$asymptotic (retarded) time $v= v_+ - v_-$ it takes $\,\!$for the transition to $\,\!$happen; the minimal external $\,\!$radius $r_+$ for which the $\,\!$classical theory can still be trusted; the minimal internal $\,\!$radius $r_-$ (it is important $\,\!$to stress that  $r$ is a temporal coordinate in the $\,\!$interior region of the black hole) reached by the black hole $\,\!$interior in region $\mathscr A$.

Each set of data $\big( m,\, v,\, r_\pm \big)$ corresponds to a different outcome of the quantum transition taking place in $\,\!$region $\mathscr B$. This means that each set of data $\big( m,\, v,\, r_\pm \big)$ uniquely defines a different quantum boundary state $\Psi \big( m,\, v,\, r_\pm \big)$ for the transition. Given the boundary state $\Psi$, any sufficiently developed tentative theory of quantum gravity should be able to assign a transition amplitude $W\big( m,\, v,\, r_\pm \big)\equiv W \Big[ \Psi \big( m,\, v,\, r_\pm \big) \Big]$ to it. The transition amplitude for the phenomenon can then be used to analyze the physics of $\,\!$region $\mathscr B$.

The next section is devoted to the computation of the transition amplitude $W\big( m,\, v,\, r_\pm \big)$ for the black-to-white hole transition using the spin foam approach.

\section{Spin foam framework}

The spin foam formalism\cite{book:Rovelli_Vidotto_CovariantLQG,article:Perez_2013} is a tentative path integral quantization of general relativity. The current $\,\!$state of the art is the $\,\!$EPRL-KKL spin foam model.\cite{EPRL,KKL,2011_generalized_spinfoams} The theory is regularized and concretely defined in the discrete setting. The boundary $\Sigma$ of a generic quantum $\,\!$region is discretized by a graph $\Gamma$ with a $\,\!$finite number of nodes and a boundary Hilbert $\,\!$space $\mathcal{H}_{\Gamma}$ is assigned to it. The latter $\,\!$is the space of $\mathrm{SU}(2)$ spin-networks on $\Gamma$. Namely, the boundary Hilbert space is $\mathcal{H}_{\Gamma}:=L^2 \big[ \SU^L   / \SU^N \big]_{\Gamma}  $, where $N $ and $\,\! L$ are respectively the total $\,\!$number of nodes and $\,\!$links in $\Gamma$. A boundary state $\,\!$is then given by a square integrable $\,\!$function $\psi (   \{ h_{\ell} \})$ that is gauge $\,\!$invariant at every node $\n \in \Gamma$. Each $h_{\ell}   \in \SU$ can $\,\!$be seen as the holonomy of the $\,\!$Ashtekar-Barbero connection along the link $\ell \in \Gamma$. The interior of the quantum region is then discretized by a $\,\!$two-complex $\mathcal C$ ($\partial\mathcal C \equiv \Gamma$) with a finite number of vertices.

Let $\f$, $\e$, $\v \in \mathcal C$ denote $\,\!$respectively a $\,\!$face, an edge and a $\,\!$vertex in $\mathcal C$. To each internal $\,\!$oriented edge $\mathcal C \ni \e:= (\v , \v ') $ having source in $\v$ and target in $\v '$ are assigned two $\SL$ $\,\!$elements $g_{\v\e}=g_{\e\v}^{-1}$ and $g_{\e\v '}=g_{\v ' \e}^{-1}$. The oriented $\,\!$edge $\e^{-1}:=( \v ' , \v)$ is the $\,\!$edge $\e$ with opposite orientation. To each boundary $\,\!$edge $\E\in \mathcal C$, that is an $\,\!$edge linking an internal vertex $\v \in \mathcal C$ and a $\,\!$node $\n \in \Gamma$, is assigned an $\SL$ element $g_{\v\n}=g_{\n\v}^{-1}$. A face $\f:=(\e_1 , \, \dots\, , \e_n )$ is the oriented $\,\!$face bounded by the $\,\!$edges $\e_1 , \, \dots\, , \e_n $. The orientation of the face $\,\!$is given by the ordering $\,\!$of the edges. For easier reading, the latters are written oriented $\,\!$accordingly to the orientation that the face induces on $\,\!$them. An internal $\,\!$face $\f \in \mathcal{B} := \mathcal{C} / \Gamma$ is a $\,\!$face bounded by vertices $\,\!$and internal edges. A boundary face $\F \in \Gamma$ is a $\,\!$face containing a $\,\!$link $\ell \in \Gamma$ in its $\,\!$boundary. Finally, a $\,\!$face $\f \ni \e$ is a face $\f$ containing $\,\!$the edge $\e$ in its $\,\!$boundary, a face $\f \ni \v$ is a face $\f$ containing the $\,\!$vertex $\v$ in its boundary and an $\,\!$edge $\e \ni \v$ is an edge $\e$ containing the $\,\!$vertex $\v$ in its boundary.

Given an arbitrary $\,\!$boundary state $\psi \in \mathcal{H}_{\Gamma}$, the theory $\,\!$assigns to it the amplitude 
\begin{equation}
W_{\mathcal{C}} \,\!\big[ \psi \big] = \,\! \int_{\SU} \prod_{\ell \in \Gamma} \,\! \diff h_{\ell} \,\!\: W_{\mathcal{C}} \,\!\big( \{h_{\ell} \} \big)\: \,\!\psi \big ( \{h_{\ell} \} \big) \,\! \, ,
\label{eq:sf1}
\end{equation}
where the $\,\!$two-complex amplitude $W_{\mathcal{C}} ( \{h_{\ell} \})$ can be defined in $\,\!$terms of elementary face amplitudes as
\begin{equation}
 W_{\mathcal{C}}\,\! \big( \{ h_{\ell} \} \big)=\,\! \int_{\SL} 
 \bigg[ \:\,\! \prod_{\v \in \mathcal C} \prod_{\e \ni \v} {}^{'} \diff g_{\v\e}\:\,\! \bigg]
  \bigg[ \:\,\! \prod_{\f \in \mathcal B} A_{\f} \big(\{g_{\v\e}\} \big)\: \bigg] \,\!
  \bigg[ \:\,\! \prod_{\F \in \Gamma} A_{\F} \big(\{g_{\v\e}\}, \, h_{\ell_\F} \big)\: \,\!\bigg]\, .
\label{eq:sf2}
\end{equation}
The product $  \prod_{\e     \ni \v} {}^{'} $ stands for the product $\,\!$over all the edges $\e \ni \v$ except $\,\!$one (which can be chosen $\,\!$arbitrarily at each vertex), $\ell_\F$ is the $\,\!$unique link in $\Gamma$ that $\,\!$belongs to the boundary $\,\!$face $\F \in \Gamma$ and the face amplitudes $A_{\f}(\{g_{\v\e}\})$ and $A_{\F} (\{g_{\v\e}\}, \, h_{\ell_\F} )$ are $\,\!$given by\footnote{For the sake of simplicity, the $\,\!$orientation of every edge in each face is assumed to $\,\!$be the one induced from the face orientation. Since $\,\!$however this kind of orientation cannot be $\,\!$implemented consistently throughout the whole $\,\!$two-complex, in actual computations the orientation $\,\!$of the edges needs to be taken into account properly.} 
\begin{equation}
A_{\f} \,\!\big( \{g_{\v\e}\} \big) := \,\! \sum_{j_\f} d_{j_\f} \mathrm{Tr} \Big[ D_{\gamma}^{(j_\f)} \,\! \big( g_{\e\v} g_{\v\e '}\big)\cdots D_{\gamma}^{(j_\f)} \,\! \big( g_{\e^{(n)}\v^{n}} g_{\v^{(n)}\e}\big) \Big] \,\! \, ,
\label{eq:sf3}
\end{equation}
\begin{equation}
A_{\F} \,\! \big( \{g_{\v\e}\},\, h_{\ell_\F} \big) := \,\! \sum_{j_\F} d_{j_\F} \mathrm{Tr} \Big[ D_{\gamma}^{(j_\F)} \big( g_{\n_t \v} g_{\v\e '}\big)\,\! \cdots \,\! D_{\gamma}^{(j_\F)} \big( g_{\e^{(n)}\v^{n}} g_{\v^{(n)} \n_s}\big)\,\!  D^{(j_\F)} \big( h_{\ell_\F} \big) \Big] \,\! \, .
\label{eq:sf4}
\end{equation}
The $\,\!$matrix $  D^{ (j )  } $ is $\,\!$the Wigner D-matrix of $\,\!$the $d_j$-dimensional $\,\!$($d_j=2j+1$) representation $\,\!$of $\SU$. The matrix $D_{\gamma}^{(j)}$ is $\,\!$the $d_j  \,\!    \!  \times   \,\!     d_j$ matrix $\big[D_{\gamma}^{(j)}\big]_{mn}(g)~=~D^{(\gamma j, j)}_{jm\;jn}(g)$, $g\in \SL$, where $D^{(\rho,k)}_{lm\: jn}$ are the matrix $\,\!$elements of the $(\rho,k)$ unitary $\,\!$representation of the principal $\,\!$series of $\SL$ in the basis $\,\!$labeled by the eigenfunctions of $L^2$ and $L_z$. The nodes $\n_s$ and $\n_t$ are $\,\!$respectively the source and the $\,\!$target of the link $\ell_\F$ and $\gamma$ is the $\,\!$Barbero-Immirzi parameter. The value $\,\!$of the label $(n)$ in $\v^{(n)}$ and $\e^{(n)}$ is $\,\!$fixed for each face by $\,\!$the topology of the two-complex $\mathcal C$.

\subsection{Discretization of region $\mathscr B$}

In order to exploit the spin foam formalism to analyze the physics of $\,\!$region $\mathscr B$, the latter needs to be properly discretized. There is no $\,\!$unique or right way to perform the $\,\!$discretization. However, in order to get simpler and clearer calculations throughout the spin foam analysis it is particularly useful to preserve as much symmetries as possible during the procedure. In this subsection the combinatorial definition of both the cellular decomposition discretizing region $\mathscr B$ and its dual two-complex are given. Although the combinatorial definition of these objects is compact and complete, it does not convey as much geometrical insight as graphical representations do. For this reason, the interested reader is strongly encouraged to look at the graphical representations reported in Refs.~\citenum{article:DAmbrosio_2021} and~\citenum{article:soltani2021}.

As it can be seen from Fig.~\ref{fig:sigma}, the topology of $\,\!$region $\mathscr B$ is $S^2 \times [0,1] \times [0,1]$ and the $\,\!$topology of $\Sigma$ is $S^2 \times S^1$. The geometry of $\Sigma$ has two symmetries: the spherical $\mathrm{SO}(3)$ $\,\!$symmetry and the $Z_2$ $\,\!$time-reversal symmetry that $\,\!$exchanges $p$ and $f$. There is also an additional topological $Z_2$ symmetry $\,\!$that exchanges $+$ and $-$. Since however the $\,\!$geometry of $S_+$ and $S_-$ is $\,\!$different, this is not a symmetry of the geometry $\,\!$of $\Sigma$. Although the topology of $\Sigma$ is not easy to$\,\!$ discretize while preserving its $\,\!$symmetries, the discretization presented in this subsection is able to accomplish this task.

Before introducing the discretization, it is useful to fix some notational conventions. Let $a,\,b,\,c,\,d$ be $\,\!$indices taking $\,\!$values in the $\,\!$set $\{1, \,\! 2 ,\,\!   3, \,\!4    \}$, $t$ be $\,\!$an index taking $\,\!$values in the $\,\!$set $\{p   , \,\! f    \}$ and $\epsilon$ be $\,\!$an index taking $\,\!$values in $\,\!$the set $\{    + ,   \,\! -\}$. If the same $\,\!$expression contains several $\,\!$indices $a,\,b,\,c,\,d$, they are$\,\!$ assumed to be all $\,\!$different from each other. The order $\,\!$of two consecutive indices $\,\!$is not important and the $\,\!$exchange of these indices results in the same element. If however $\,\!$two indices are separated by a $\,\!$comma, the exchange of these $\,\!$indices results in a different $\,\!$element.

Let $p_a^+$ be four $\,\!$points on $S_+$ and $p_a^-$ be $\,\!$four points on $S_-$. The three-dimensional $\,\!$triangulation discretizing $\Sigma$ is $\,\!$then defined by the $\,\!$points $p^{\epsilon}_a$, the $\,\!$segments \(s_{ab}^{\epsilon}\) and $\,\!$\(s_{a, \, \! b}^t\), the $\,\!$triangles \(L_{a}^{\epsilon}\) and \(L_{a,\,\! b}^{t   \epsilon}\), the $\,\!$tetrahedra \( N_{a}^{t   \epsilon}\) and \( N_{ab}^{t}\), and $\,\!$their boundary $\,\!$relations:
\begin{align}
&\partial \,\! s_{a \,\!b}^{\epsilon} =(p_{a}^{\epsilon} , p_{b}^{\epsilon} ); \,\!\\
&\partial \,\! s_{a, \,\!b}^{t}=(p_{a}^{-},p_{b}^{+})^{t};\,\!\\
&\partial \,\! L_{a}^{\epsilon}=(s_{b \,\! c}^{\epsilon},s_{c \,\!d}^{\epsilon},s_{d \,\! b}^{\epsilon});\,\!\\
&\partial \,\! L^{t \,\! +}_{a, \,\! b}=(s^{+}_{cd},s^{t}_{a,\,\! c},s^t_{a,\,\! d}); \,\!\\
&\partial \,\! L^{t \,\! -}_{a, \,\! b}=(s^{-}_{c \,\! d},s^{t}_{c, \,\! a},s^{t}_{d, \,\! a}); \,\!\\
&\partial \,\! N^{t\,\! \epsilon}_{a}=(L^{t \,\! \epsilon}_{a, \,\! b},L^{t \,\! \epsilon}_{a, \,\! c},L^{t \,\! \epsilon}_{a, \,\! d},L_{a}^{\epsilon} ); \,\!\\
&\partial \,\! N^{t}_{a \,\! b}=(L^{t \,\! +}_{a, \,\! b},L^{t \,\! +}_{b, \,\! a},L^{t \,\! -}_{c, \,\! d},L^{t \,\! -}_{d, \,\! c}).\,\!
\end{align}
Besides being a three-dimensional $\,\!$object of its own, this triangulation $\,\!$serves as the boundary of the cellular decomposition discretizing $\,\!$region $\mathscr B$. This four-dimensional $\,\!$cellular decomposition is defined by $\,\!$the two-dimensional surfaces \(f_{a, \,\! b}\), the three-dimensional $\,\!$cells \(e_{a \,\! ,b}^{\epsilon}\), the four-dimensional $\,\!$cells \(v_{a}^{\epsilon}\) and \(v_{a \,\! b}\), and their $\,\!$boundary relations:
\begin{align}
&\partial \,\!f_{a \,\! ,b}=(s^{p}_{a \,\! ,b},s^{f}_{a \,\! ,b}); \,\!\\
&\partial \;\! e_{a, \,\! b}^{\epsilon}=(L^{p \,\! \epsilon}_{a, \,\! b},L^{f \,\! \epsilon}_{a \,\! ,b}, f_{a, \,\! c}, f_{a \,\! ,d});\,\!\\
&\partial \;\! v_{a}^{\epsilon}=(N^{p \,\! \epsilon}_{a} ,N_{a}^{f \,\! \epsilon},e_{a \,\! ,b}^{\epsilon},e_{a \,\! ,c}^{\epsilon},e_{a, \,\! d}^{\epsilon} );\,\!\\
&\partial \;\! v_{a \,\! b}=(N^{p}_{a \,\! b},N^{f}_{a \,\! b},e^{+}_{a \,\! ,b},e^{+}_{b \,\! ,a},e^{-}_{c \,\! ,d},e^{-}_{d \,\! ,c}).\,\!
\end{align}
Note that this cellular $\,\!$decomposition is not a $\,\!$triangulation.

The spin foam formalism is defined using the $\,\!$discrete object dual to the cellular decomposition. The graph $\Gamma$ $\,\!$dual to the three-dimensional $\,\!$triangulation of $\Sigma$ is defined $\,\!$as
\begin{align}
\mathrm{Nodes}\textcolor{white}{aaaaaaaaaaaaaaa} 
\n^{t \, \!\epsilon}_{a} \textcolor{white}{a} \mathrm{and} \textcolor{white}{a} \n^{t}_{a \, \!b}; 
\textcolor{white}{aaaaaaaaaaaaaaaaaaaaaaaa.aaa}\\
\mathrm{Links}\textcolor{white}{aaaaaaaaaaaaaaa}
\ell_{a}^{\epsilon}= (\n^{p \epsilon}_ {a} , \n^{f \epsilon}_{a}); 
\textcolor{white}{aaaaaaaaaaaaaaaaaaaaaaaaaa}\\
\ell^{t \,\! +}_{a, \,\!b}=(\n^{t}_{a \,\! b},\n^{t \,\! +}_{a}); 
\textcolor{white}{aaaaaaaaa.aaaaaaaaaaaaaaa}\\
\ell^{t \,\! -}_{a \,\! ,b}=(\n^{t}_{c \,\! d},\n^{t \,\! -}_{a}).
\textcolor{white}{aaaaaaaaaaaaaaaaaaaaaaaaa}
\end{align}
The $\,\!$two-complex $\mathcal{C}$, $\,\!$whose boundary $\partial \!\:\,\! \mathcal{C}$ is $\Gamma$, dual $\,\!$to the four-dimensional cellular $\,\!$decomposition of $\,\!$region $\mathscr B$ is defined $\,\!$as
\begin{align}
\mathrm{Vertices}\textcolor{white}{aaaaa..aaaaaaaa} 
\v^{\epsilon}_{a} \textcolor{white}{a} \mathrm{and} \textcolor{white}{a} \v_{a \,\! b}; 
\textcolor{white}{aaaaaaaaaaaaaaaaaaaaaaaaaaaaa}\\
\mathrm{Edges}\textcolor{white}{aaa..aaaaaaaaaaa}
\E^{t \,\! \epsilon}_{a} = (\v^{\epsilon}_{a},\n^{t\,\!\epsilon}_{a}); 
\textcolor{white}{aaaaaaaaaaaaaaaaaaaaaaaaaaa}\\
\E^{t}_{a\,\! b}=(\v_{a\,\! b},\n^{t}_{a\,\! b}); 
\textcolor{white}{aaaaaaaaaaaaaaaa.aaaaaaaaa}\\
\e^{+}_{a,\,\! b}=(\v^{+}_{a},\v_{a\,\! b});
\textcolor{white}{aaaaaaaalaaaaaaaaaa..aaaaaa}\\
\e^{-}_{a,\,\! b}=(\v^{-}_{a},\v_{c\,\! d});
\textcolor{white}{aaaaaaaaaalaaaaaaaaaa..aaaa}\\
\mathrm{Faces}\textcolor{white}{aaaa.aaaa.aaaaaa}
\F^{\epsilon}_{a}=\big(\ell^{\epsilon}_{a},(\E^{f \,\! \epsilon}_{a})^{-1},\E^{p \,\! \epsilon}_{a}\big); 
\textcolor{white}{aaaaaaaaaaaaaaa.aaaa}\\
\F^{t \,\!+}_{a, \,\! b}=\big(\ell^{t \,\!+}_{a\,\! ,b},
(\E^{t \,\! +}_{a})^{-1},\e^{+}_{a,\,\! b}, \E^{t}_{a\,\! b}\big);
\textcolor{white}{aaaaaaaaaaaaa}\\
\F^{t \,\! -}_{a\,\! ,b}=\big(\ell^{t \,\! -}_{a\,\! ,b},
(\E^{t \,\! -}_{a})^{-1},\E^{-}_{a\,\! ,b},
\E^{t}_{c\,\! d}\big);
\textcolor{white}{aaa.aaaaaaaaa}\\
\f_{a,\,\! b} \overset{c\,\! <d}{=} \big( \e^{+}_{a\,\! ,c},(\e^{-}_{b\,\! ,d} )^{-1},\e^{-}_{b\,\!,c},(\e^{+}_{a\,\! ,d})^{-1} \big).
\textcolor{white}{.aaaaaaaaa}
\end{align}
The orientation of each $\,\!$element of the two-complex $\,\!$can be easily read from $\,\!$this combinatorial definition.

It is interesting to $\,\!$analyze how many of the topological symmetries of $\,\!$region $\mathscr{B}$ are preserved under $\,\!$this discretization. The $\,\!$spherical $\mathrm{SO}(3)$ symmetry $\,\!$of region $\mathscr B$ is discretized to a $\,\!$tetrahedral symmetry of the $\,\!$two-complex, which $\,\!$is realized as an even $\,\!$permutation of the indices $a,\,\! b \,\! , \,\! c, \,\! d$. The $Z_2$ time-reversal $\,\!$symmetry that exchanges the indices $p$ and $f$ and $\,\!$the $Z_2$ symmetry that $\,\!$exchanges the indices $+$ and $-$ are instead preserved $\,\!$exactly.

\subsection{Discrete geometrical data and boundary state}

The continuous geometry $\,\!$of $\Sigma$ is approximated by $\,\!$the assignment of discrete geometrical $\,\!$data to the triangulation discretizing $\Sigma$ (or equivalently to is dual graph $\Gamma$). There $\,\!$is once again no $\,\!$unique or right way to $\,\!$do it. Different assignments $\,\!$of discrete $\,\!$geometrical data $\,\!$simply define different $\,\!$approximations of the same continuous geometry. The discrete geometry presented $\,\!$in this subsection preserves the two geometrical $\,\!$symmetries of the continuous geometry of $\Sigma$.

To preserve the $Z_2$ time-reversal symmetry is sufficient to firstly construct the discrete geometrical data for the triangulation discretizing $\Sigma^p$ and then to define the discrete geometrical data for the triangulation discretizing $\Sigma^f$ as their time reversal. The sphere $S_{+}$ ($S_{-}$) in $\Sigma$ is discretized by the tetrahedron bounded by the four triangles $\{ L^{+}_{a} \}$ ($\{ L^{-}_{a}\}$) in the triangulation. In order to preserve as much symmetry as possible in the discretization process of the geometry, the same area $a_+=\pi r_{+}^2$ ($a_-=\pi r_{-}^2$), which is one fourth of the area of the sphere $S_{+}$ ($S_{-}$), is assigned to each triangle $L^{+}_{a}$ ($L^{-}_{a}$).

The continuous $\,\!$intrinsic geometry of $\Sigma^p$ is specified $\,\!$by the line element
\begin{equation}
\diff s^2 \,\!= f^2 (r) \diff r^2 + \,\! r^2 \diff \Omega^2\,\!\, ,
\label{eq:linelement1}
\end{equation}
in which $f^2(r)\,\!=\beta \, \big( 2- \beta \,\!\, ( 1 - 2m/r ) \big)\,\!$ on $\Sigma^{p}_{+}$ $\,\!$and $f^2(r)=1$ on $\Sigma^{p}_{-}$. To begin with, this line element is approximated with
\begin{equation}
\diff s^2 =\,\! \xi^2 \diff r^2 + \,\! r^2 \diff \Omega^2\, ,
\label{eq:linelement2}
\end{equation}
where $\xi$ is a constant that is fixed by $\,\!$requiring the volume of $\Sigma^p$ computed with the $\,\!$line element in Eq.~(\ref{eq:linelement2}) to coincide with the $\,\!$volume of $\Sigma^p$ computed with the line element in Eq.~(\ref{eq:linelement1}). This condition uniquely $\,\!$fixes the value of $\xi$ in terms of the four $\,\!$parameters $\big( m,\, v,\, r_\pm \big)$ characterizing the continuous geometry of $\Sigma^p$. Requiring the $\,\!$discrete geometrical data to preserve the topological $\,\!$tetrahedral symmetry of the triangulation of $\Sigma$, the same area $A_+$ ($A_-$) must be assigned to $\,\!$each triangle $L^{p +}_{a,b}$ ($L^{p -}_{a,b}$). The line element in Eq.~(\ref{eq:linelement2}) is then used to $\,\!$compute the value of $A_\pm$, which is
\begin{equation}
A_{\pm} = \,\! \pi \,\! r_{\pm}^2 \: \sqrt{ \, \frac{\,\xi^2}{\,\! 18}\,\! 
\Big( 1- \,\! 3 \,\frac{r_{\mp}}{r_{\pm}}\Big)^{\,\! 2} + \frac{\,\! 2}{3}}\,\!\, .
\end{equation}
The symmetry conditions together with the values of the areas $a_\pm$ and $A_\pm$ completely define the discrete intrinsic geometry of the triangulation of $\Sigma$. Analogously, the discrete extrinsic geometry of the triangulation is completely defined by the assignment of the extrinsic angles $\theta_\pm$ between $N^{p\pm}_{a}$ and $N^{f\pm}_{a}$ at $L^{\pm}_{a}$ and the extrinsic angles $\Theta_\pm$ between $N^{p\pm}_{a}$ and $N^{p}_{a,b}$ at $L^{p\pm}_{a,b}$.

Let $n_{\mu}^{t \epsilon}$ be the normal one-form of $\Sigma^{t}_{\epsilon}$. From the definition of $\theta_\pm$ it follows that
\begin{equation}
\cos \,\!(\theta_\pm) =\,\! \Big(\, g^{\mu\nu} \, n_{\mu}^{f\, \pm} \,\!\,
n_{\nu}^{p\, \pm}\, \Big)\Big|_{S_\pm}\,\! \, ,
\end{equation}
where $g^{\mu\nu}$ is the inverse of the metric tensor defined by the line element in Eq.~(\ref{eq:linelement1}), giving
\begin{equation}
\cos (\theta_{+} )\,\! = \frac{1+\Big[ \big( 1- 2m/r_{+} \big) \,\!\, \beta - \,\! 1 
\Big]^2 \,\!}{\:\big|\, \beta \, \big(\,\beta - 2 -2m\beta / r_{+} \big) \,\!\big|
\, \big( 1 - \,\! 2m / r_{+} \big)}
\end{equation}
and
\begin{equation}
\cos \,\!(\theta_{-})= \,\!\frac{1 \,\! +2m/r_{-}}{\,\! 1 -\,\! 2m/r_{-}}\,\!\, .
\end{equation}
The angles $\Theta_\pm$ represent a discrete approximation of the continuous extrinsic curvature of $\Sigma_{\pm}$. A convenient discretization that preserves the symmetries of the triangulation is
\begin{equation}
\Theta_\pm = \frac{1}{12} \int_{\Sigma^{p}_{\pm}} \big( k^\pm \big)^{i}_{\textcolor{white}{i.}i}\, ,
\end{equation}
where $k^{\pm}_{ij}$ is the extrinsic curvature tensor of $\Sigma^{p}_{\pm}$ defined in Eqs.~(\ref{eq:k1}) and~(\ref{eq:k2}).

The discrete geometry of the triangulation of $\Sigma$ has thus been explicitly constructed in terms of the four parameters $\big( m,\, v,\, r_\pm \big)$ characterizing the continuous geometry of $\Sigma$. From the point of view of the dual graph $\Gamma$, the discrete geometry consists in the assignment of an area and an angle to each link of $\Gamma$. The area assigned to each link represents the area of the triangle that is dual to the link and the angle assigned to each link represents the extrinsic curvature between the two tetrahedra that share the triangle dual to the link. This geometrical data uniquely specifies an extrinsic coherent state\cite{book:Rovelli_Vidotto_CovariantLQG} $\Psi_{BW}= \Psi_{BW} \big( a_\pm , \, \theta_\pm , \, A_\pm , \, \Theta_\pm \big) =\Psi_{BW} \big( m,\, v,\, r_\pm \big) \in \mathcal{H}_{\Gamma}$ that is peaked on the discrete classical geometry defined by $\big( a_\pm , \, \theta_\pm , \, A_\pm , \, \Theta_\pm \big)$. The boundary state $\Psi_{BW}\in \mathcal{H}_{\Gamma}$ is the quantum state representing the outcome of the transition taking place in $\,\!$region $\mathscr B$.

\subsection{Transition amplitude}

Having explicitly $\,\!$constructed the two-complex discretizing $\,\!$region $\mathscr B$ and the boundary $\,\!$state describing the outcome of the quantum $\,\!$transition, the $\,\!$spin foam transition amplitude for $\,\!$the black-to-white $\,\!$hole transition can be readily $\,\!$computed using the formulas in Eqs.~(\ref{eq:sf1}) and~(\ref{eq:sf2}).

The assignment of group elements to the edges and the links of the two-complex $\mathcal{C}$ is:
\begin{flalign}
& \textcolor{white}{aaaaaaaaaaaaaaaaaaaaa} \ell^{\epsilon}_{a} \,\!\longleftrightarrow \,\! h^{\epsilon}_{a} \in \SU \,\,\! ;&&\\
& \textcolor{white}{aaaaaaaaaaaaaaaaaaaaa}  \ell^{t \,\! \epsilon}_{a \,\! ,b} \,\! \longleftrightarrow \,\! h^{ t \,\!\epsilon}_{a,\,\! b} \in \,\! \SU\,\,\! ;&&\\
& \textcolor{white}{aaaaaaaaaaaaaaaaaaaaa} \E^{t \,\! \epsilon}_{a} \longleftrightarrow \,\! g^{t \,\! \epsilon}_{a} \,\! \in \SL\, ;&&\\
& \textcolor{white}{aaaaaaaaaaaaaaaaaaaaa} \E^{t}_{a\,\! b} \,\! \longleftrightarrow  \,\! g^{t}_{a \,\! b} \in \SL \,\,\! ;&&\\
& \textcolor{white}{aaaaaaaaaaaaaaaaaaaaa} \e^{\epsilon}_{a\,\! ,b} \,\! \longleftrightarrow g^{\epsilon}_{a \,\! \rightarrow \,\! b}\,\,\! , \,\!\, g^{\epsilon}_{a\,\! \leftarrow \,\! b}\in \SL\, \,\! . &&
\end{flalign}
The group $\,\!$elements $g^{\epsilon}_{a \,\! \leftrightarrow b}$ are assigned to $\,\!$the two oriented $\,\!$half-edges of $\mathrm{e}^{\epsilon}_{a\,\! ,b}$. The $\,\!$element $g^{\epsilon}_{a \,\! \rightarrow b}$ is assigned $\,\!$to the oriented half-edge $\,\!$with source in the $\,\!$source of $\mathrm{e}^{\epsilon}_{a,\,\! b}$ and target in $\,\!$the center of $\mathrm{e}^{\epsilon}_{a\,\! ,b}$. The element $g^{\epsilon}_{a \,\! \leftarrow b}$ $\,\!$is assigned to the oriented $\,\!$half-edge with source $\,\!$in the target of $\mathrm{e}^{\epsilon}_{a,\,\! b}$ and target $\,\!$in the center of $\mathrm{e}^{\epsilon}_{a\,\! ,b}$. Carefully considering the topology and the orientation $\,\!$pattern of the $\,\!$two-complex, the elementary face $\,\!$amplitudes (see Eqs.~(\ref{eq:sf3}) and~(\ref{eq:sf4})$\,$) for $\mathcal C$ can be computed:
\begin{equation}
A^{\epsilon}_{a} \,\! \big( \{g^{t \,\! \epsilon}_{a}\} \,\!, \, h^{\epsilon}_{a} \, \big) \,\!=
\displaystyle \sum_{j} \,\! d_j \, \mathrm{Tr} \,\! \Big[D^{(j)}_{\gamma} \,\!\big(\, \big( \, g^{f \,\!\epsilon}_{a} \, \big)^{-1} \: g^{p \,\! \epsilon}_{a} \, \big) \, D^{(j)}\big( h^{\epsilon}_{a} \big)\,\! \Big]\, ,
\label{eq:fa1}
\end{equation}
\begin{equation}
\begin{split}
A^{t \,\! +}_{a\,\! ,b} \,\! \big( \,g^{t\,\! +}_a,\, g^{t}_{a\,\! b}, \,\!\, g^{+}_{a\leftrightarrow \,\! b},\,\!  h^{t \,\! +}_{a,\,\! b}\, \big) \,\!
= \displaystyle\,\! \sum_{j} d_j\,  \,\!\mathrm{Tr} \Big[\,
& D^{(j)}_{\gamma} \,\!\big( \, \big( \, g^{t\,\! +}_{a}\, \big)^{-1} \: \,\! g^{+}_{a\rightarrow \,\!b} \,\big)  
D^{(j)}_{\gamma} \,\!\big( \, \big( \, g^{+}_{a\leftarrow \,\! b}\, \big)^{-1} \:\: g^{t}_{a\,\! b} \,  \big) \,\! \\
\times & D^{(j)} \,\!\big( \,h^{t\,\! +}_{a,\,\!b}  \,  \,\! \big) \,\Big] \, ,
\end{split}
\label{eq:fa2}
\end{equation}
\begin{equation}
\begin{split}
A^{t\,\! -}_{a\,\! ,b} \,\! \big(\, g^{t\,\!-}_{a}, \,\! g^{t}_{c\,\!d},g^{-}_{a\leftrightarrow \,\!b}, \,\! h^{t-}_{a,b}\, \big)
= \displaystyle\sum_{j} \,\! d_j\,  \mathrm{Tr} \Big[\,
& D^{(j)}_{\gamma} \,\! \big( \, \big( \, g^{t\,\!-}_{a} \, \big)^{-1} \:\,\!  g^{-}_{a\rightarrow \,\!b} \,\big)  
D^{(j)}_{\gamma} \,\! \big( \, \big( \, g^{-}_{a\,\! \leftarrow b}\, \big)^{-1} \:\: g^{t}_{c\,\!d}   \,\big) \,\! \\ 
\times & D^{(j)}  \,\! \big( \, h^{t\,\!-}_{a,\,\!b}\, \,\! \big) \,\Big] \, ,  
\end{split}
\label{eq:fa3}
\end{equation}
\begin{equation}
\begin{split}
A_{a,\,\!b} \,\!  \big( \, \{g^{+}_{a \,\!\leftrightarrow c} \} \,\! , \, \,\! \{ g^-_{b \,\!\leftrightarrow c} \}\, \big) \overset{c<d}{=} \displaystyle \,\! \sum_{j}  d_{j}\, \,\! \mathrm{Tr} 
\Big[ \, & D^{(j)}_{\gamma} \,\!\big( \, \big( \, g^{+}_{a \,\! \rightarrow d} \, \big)^{-1} \,\!\: g^{+}_{a \,\! \rightarrow c} \,\big) \,
D^{(j)}_{\gamma} \,\!\big( \, \big( \, g^{+}_{a \,\! \leftarrow c}\, \big)^{-1} \,\!\: g^{-}_{b\,\! \leftarrow d} \,\big) \\
  \times \: &
D^{(j)}_{\gamma} \,\!\big( \, \big( \, g^{-}_{b\,\! \rightarrow d} \, \big)^{-1} \,\!\: g^{-}_{b\,\! \rightarrow c} \, \big) \,
D^{(j)}_{\gamma} \,\!\big( \, \big( \, g^{-}_{b\,\! \leftarrow c}\, \big)^{-1} \,\!\: g^{+}_{a\,\! \leftarrow d} \,\big)\,\Big]
\, .
\end{split}
\label{eq:fa4}
\end{equation}
The two-complex transition amplitude $W_{\mathcal{C}} ( \{h_{\ell} \})$ can then be written as
\begin{equation}
\begin{split}
 W_{\mathcal{C}} \,\! \big( \, \{ h^{\epsilon}_{a} \}, \, \{ h^{t \,\!\epsilon}_{a \,\!,b} \}\, \big)  \,\!
 = & \,\! \int_{\mathrm{SL}(2,\mathbb{C})}  
\prod_{\epsilon \,\!a} \,\!\diff g^{p \,\! \epsilon}_{a} \: \,\!
\prod_{a< \,\!b} \diff g^{p}_{a \,\! b} \:  
\prod_{\epsilon \,\! ab} \diff g^{\epsilon}_{a\leftrightarrow \,\! b}  \,\!\\
& \times 
\prod_{\epsilon \,\! a}  \,\!  A^{\epsilon}_{a} \big( \{g^{t \,\! \epsilon}_{a}\} , \, h^{\epsilon}_{a} \, \big)   
\prod_{a \,\! b} A_{a,\,\!b}  \,\!\big( \, \{g^{+}_{a \,\!\leftrightarrow c} \} \,\! , \, \,\! \{ g^-_{b \,\!\leftrightarrow c} \}\,\,\! \big)   \\ 
 & \times \prod_{t \,\!ab}\,\!  A^{t \,\! +}_{a\,\! ,b} \big( \,g^{t\,\! +}_a,\, g^{t}_{a\,\! b}, \, g^{+}_{a\leftrightarrow \,\! b}, h^{t \,\! +}_{a,\,\! b}\, \big)\,\!\\
 & \times \prod_{t \,\! ab}\,\!  A^{t\,\! -}_{a\,\! ,b} \big(\, g^{t\,\!-}_{a}, g^{t}_{c\,\!d},g^{-}_{a\leftrightarrow \,\!b},\,\! h^{t-}_{a,b}\, \big)  \, .
\end{split}
\label{eq:amplitudeC}
\end{equation}
Since Eq.~(\ref{eq:sf2}) dictates to $\,\!$drop one integration per $\,\!$vertex, the integration over the $\{g^{f\,\!\epsilon}_{a}\}$ and $\{g^{f}_{a\,\!b}\}$ variables $\,\!$has been dropped.

Finally, the black-to-white hole transition amplitude $W\big( m,\, v,\, r_\pm \big)$ is
\begin{equation}
\begin{split}
W\big( m,\, v,\, r_\pm \big) \,\!=  W \big[ \Psi_{BW}\big]= \int_{\mathrm{SU}(2) \,\!} 
 \prod_{\epsilon \,\!a} \,\! & \diff h^{\epsilon}_{a} \:
 \prod_{t \,\! \epsilon \,\!ab} \,\! \diff h^{t \,\!\epsilon}_{a,\,\!b} \;\:
W_{\mathcal{C}} \,\!\big( \{h^{\epsilon}_{a} \} , \, \,\!\{h^{t\,\!\epsilon}_{a,\,\!b} \} \big)   \\
& \times
\Psi_{\mathrm{BW}}\,\! \big( \{h^{\epsilon}_{a} \} , \, \{h^{t\,\!\epsilon}_{a,\,\!b} \} \big)\, .
\end{split}
\label{eq:amplitudeB2W}
\end{equation}
This expression contains the whole physics of the phenomenon. Its investigation is currently ongoing and it will be reported elsewhere.

\section{Summary and outlook}

The quantum region of a black hole spacetime can be divided in three different subregions: region $\mathscr{C}$, where the collapsing matter enters its quantum gravity regime; region $\mathscr B$, where the horizon reaches Planckian size at the end of the evaporation process; region $\mathscr A$, where the curvature reaches Planckian scale independently from region $\mathscr B$ and region $\mathscr C$. The principle of locality demands that these regions are independent from each other and that they can subsequently be studied separately. The evidence in favor of the black-to-white hole transition scenario resulting from the separate analysis of the physics of these regions is overwhelming.

Focusing on the black-to-white hole horizon transition in region $\mathscr B$, that is the last stage of the life of a black hole in this scenario, the physics of the boundary of the quantum region is completely determined by four parameters: the mass $m$ of the black hole at the moment in which the transition takes place; the external asymptotic (retarded) time $v$ it takes for the transition to happen; the minimal external radius $r_+$ for which the classical theory can still be trusted; the minimal internal radius $r_-$ reached by the black hole interior in region $\mathscr A$.

The spin foam formalism provides a clear framework to study this scenario. However, since the theory is properly defined in the discrete setting, to compute the transition amplitude for the phenomenon the physics of $\,\!$region $\mathscr B$ needs to be appropriately discretized. From a practical point of view the discretization needs to be both sufficiently refined to capture the relevant degrees of freedom of the phenomenon and sufficiently simple for the transition amplitude to be explicitly computed and analyzed. In this work a convenient discretization that preserves the symmetries of the continuous geometry is explicitly constructed and the resulting transition amplitude $W\big( m,\, v,\, r_\pm \big)$ is computed. Although the two-complex $\mathcal C$ discretizing region $\mathscr B$ is rather complicated, its high degree of symmetry allows the transition amplitude to be expressed in a remarkably compact way. Due to the severe complexity of spin foam amplitudes, an analytical study of the transition amplitude $W\big( m,\, v,\, r_\pm \big)$ for the black-to-white hole transition is not feasible at this point. On the other hand, recent developments\cite{Dona_sarno_numerical_methods_EPRL,Dona_numerical_study,gozzini2021highperformance,pietorpaolo} in the numerical computation of spin foam amplitudes and the high degree of symmetry of the constructed discretization should allow a numerical investigation of the transition amplitude.

In this work the black hole lifetime is assumed to be of the order of the evaporation process (although the constructed framework can describe also smaller timescales). This choice is motivated by the analysis of the black-to-white hole transition performed in Refs.~\citenum{planckstar_tunneling_time2016} and~\citenum{Characteristic_Time_Scales_B2W_2018}, where, neglecting Hawking radiation, the lifetime of the black hole was estimated to be much longer of the evaporation process, thus proving that the assumption of neglecting Hawking radiation was not justified. However, due to the use of a fairly coarse discretization and of several rough approximations it is unclear whether this result is reliable. Furthermore, recent results\cite{Kelly_2020} seems to support the black hole lifetime of the order $m_0^2$ heuristically suggested in Ref.~\citenum{Haggard_2015_fireworks}. The numerical analysis of the transition amplitude $W\big( m,\, v,\, r_\pm \big)$ computed in this work may provide an estimation of the black-to-white hole transition timescales and improve the understanding of its phenomenology.\cite{Barrau_2014,Barrau_2016,Barrau_2017,Rovelli_Vidotto2018}

\bibliographystyle{ws-procs961x669}
\bibliography{references}

\end{document}